\documentclass{article}
\usepackage{spconf,amsmath,graphicx,booktabs,array,float,tabularx}
\usepackage[hidelinks]{hyperref}
\usepackage{orcidlink}
\graphicspath{{figures/}}

\newcommand{\psnrBF}{$19.8$ }
\newcommand{\psnrBMTD}{$17.7$ }
\newcommand{\psnrFFDNet}{$38.3$ }
\newcommand{\psnrViT}{$23.5$ }
\newcommand{\psnrDAARE}[1]{\ifstrequal{#1}{b}{$\pmb{42.2}$}{$42.2$} }

\newcommand{\ssimBF}{$0.773$ }
\newcommand{\ssimBMTD}{$0.555$ }
\newcommand{\ssimFFDNet}{$0.917$ }
\newcommand{\ssimViT}{$0.817$ }
\newcommand{\ssimDAARE}[1]{\ifstrequal{#1}{b}{$\pmb{0.981}$}{$0.981$} }

\newcommand{\psnrDiff}{$3.9$ }
\newcommand{\ssimDiff}{$0.064$ }
\newcommand{\D}{\mathcal{D}}
\newcolumntype{x}[1]{>{\centering\arraybackslash\hspace{0pt}}p{#1}}

\newcommand{\ARXIV}{}
\usepackage{xcolor,mdframed,enumitem,soul}
\definecolor{dragonfruit}{rgb}{1,0,0.5}
\newcommand{\ic}[1]{
\ifdefined\DEBUG
    \begin{mdframed}
    [linecolor=dragonfruit,
     linewidth=2pt,
     topline=false,
     rightline=false,
     bottomline=false]
    \color{dragonfruit}
    #1
    \end{mdframed}
\fi
}
\newcommand{\orcid}[1]{
\ifdefined\ARXIV
    \hspace{-5pt}\orcidlink{#1}\hspace{-2pt}
\fi
}
\newcommand{\ai}[1]{
\ifdefined\DEBUG
    \begin{mdframed}
    [linecolor=gray,
     linewidth=2pt,
     topline=false,
     rightline=false,
     bottomline=false]
    \color{gray}
    #1
    \end{mdframed}
\fi
}

\title{Removing Radio Frequency Interference from Auroral Kilometric Radiation with Stacked AutoEncoders}

\name{
Allen Chang$^{1,2}$,
\orcid{0000-0002-5318-7660}Mary Knapp$^{2}$,
\orcid{0000-0002-0772-8825}James LaBelle$^{3}$,
\orcid{0000-0003-2627-2031}John Swoboda$^2$,
\orcid{0000-0002-7504-0336}Ryan Volz$^2$,
\orcid{0000-0002-0031-9324}Philip J. Erickson$^2$
\thanks{
This work was supported by National Science Foundation grant AST-1950348.}
}
\address{
$^1$ Department of Computer Science, University of Southern California, Los Angeles, USA\\
$^2$ Haystack Observatory, Massachusetts Institute of Technology, Westford, USA\\
$^3$ Department of Physics and Astronomy, Dartmouth College, Hanover, USA
}

\begin{document}
\maketitle

%

\begin{abstract}
\ic{ 1.1 What is AKR
\begin{enumerate}[leftmargin=1.5em]\itemsep -0.5em
    \item What is auroral kilometric radiation
    \item About and sources of AKR noise
    \item Our work and methods
\end{enumerate}
}
Radio frequency data in astronomy enable scientists to analyze astrophysical phenomena. However, these data can be corrupted by radio frequency interference (RFI) that limits the observation of underlying natural processes. 
In this study, we extend recent developments in deep learning algorithms to astronomy data. 
We remove RFI from time-frequency spectrograms containing auroral kilometric radiation (AKR), a coherent radio emission originating from the Earth's auroral zones that is used to study astrophysical plasmas. 
We propose a Denoising Autoencoder for Auroral Radio Emissions (DAARE) trained with synthetic spectrograms to denoise AKR signals collected at the South Pole Station. DAARE achieves \psnrDAARE{} peak signal-to-noise ratio (PSNR) and \ssimDAARE{} structural similarity (SSIM) on synthesized AKR observations, improving PSNR by \psnrDiff and SSIM by \ssimDiff compared to state-of-the-art filtering and denoising networks. Qualitative comparisons demonstrate DAARE's capability to effectively remove RFI from real AKR observations, despite being trained completely on a dataset of simulated AKR. The framework for simulating AKR, training DAARE, and employing DAARE can be accessed at \href{https://github.com/Cylumn/daare}{\texttt{github.com/Cylumn/daare}}.
\end{abstract}
\ic{Up to 5 Index Terms}

\ic{Acronyms:

\begin{enumerate}[leftmargin=1.5em]\itemsep -0.5em
    \item Auroral Kilometric Radiation (AKR)
    \item Radio Frequency Interference (RFI)
    \item Convolutional Neural Network (CNN)
    \item Convolutional Denoising AutoEncoder (CDAE)
    \item Denoising Autoencoder for Auroral Radio Emissions (DAARE)
    \item Mean Squared Error (MSE)
    \item Structural SIMilarity (SSIM)
    \item Multi-Scale Structural SIMilarity (MS-SSIM)
    \item Bilateral Filtering (BF)
    \item Block-Matching and 3D filtering (BM3D)
    \item Fast and Flexible Denoising Network (FFDNet)
    \item Vision Transformer (ViT)
    \item Peak Signal-to-Noise Ratio (PSNR)
    \item Structural SIMilarity (SSIM)
\end{enumerate}
}
\section{Introduction}\label{sec:introduction}

\ic{\ref{sec:introduction}.1 Background
\begin{enumerate}[leftmargin=1.5em]\itemsep -0.5em
    \item Radio astronomy applications
    \item What is auroral kilometric radiation
    \item Sources of AKR noise
    \item Problem: Why is AKR spectrogram denoising important
\end{enumerate}
}
Analysis of astronomical radio signals commonly utilize time-frequency spectrograms to visualize electromagnetic data across a range of frequencies with respect to time. Past work investigating the Earth's auroral region visualize radio frequency signals as spectrograms to identify auroral kilometric radiation (AKR), enabling new insight into wave-particle processes in astrophysical plasmas \cite{gurnett1974earth, labelle2011ground}. 
However, interference from local or distant transmitters can corrupt the signals of interest.
Denoising AKR signals is a critical preprocessing stage to improve downstream analysis of auroral emissions and can enhance our understanding of space physics. 

\ic{\ref{sec:introduction}.2 Related Work
\begin{enumerate}[leftmargin=1.5em]\itemsep -0.5em
    \item What are some current denoising methods
    \item Limitations of current denoising methods
\end{enumerate}
}

Spectrographic representation of AKR signals enables application of image processing algorithms for noise removal. Classical image denoising methods apply a mixture of local and non-local filtering \cite{buades2005review, dabov2007image}.
However, traditional algorithms struggle with deciphering aggressive noise structures or recovering masked features \cite{hasan2018improved}. Recent works in convolutional neural networks (CNN) and convolutional denoising autoencoders (CDAE) show that these approaches can outperform traditional algorithms in denoising and reconstruction \cite{zhang2018ffdnet}. Further, Vision Transformers \cite{dosovitskiy2020image} have demonstrated state-of-the-art ability in masked feature recovery when trained on ample data. We extend these works in a space physics context by evaluating their ability to address radio frequency interference (RFI) present in AKR spectrograms \cite{labelle2011ground}.

\ic{\ref{sec:introduction}.4 Technical Approach
\begin{enumerate}[leftmargin=1.5em]\itemsep -0.5em
    \item What architectures do we try?
    \item How do we evaluate?
    \item Contribution 1: Our method is better
    \item Contribution 2: We trained with synthetic data, and it works
\end{enumerate}
}
In this paper, we remove RFI from AKR spectrograms observed at the South Pole Station. We propose a Denoising Autoencoder for Auroral Radio Emissions (DAARE) that uses successively stacked CDAEs trained on simulated data and mean squared error (MSE) loss for noise removal.
We evaluate DAARE against bilateral filtering (BF) \cite{tomasi1998bilateral}, block-matching and 3D filtering (BM3D) \cite{dabov2007image}, the Fast and Flexible Denoising Network (FFDNet) \cite{zhang2018ffdnet}, and the Vision Transformer (ViT) \cite{dosovitskiy2020image} with two metrics: peak signal-to-noise ratio (PSNR) and structural similarity (SSIM) \cite{wang2004image}. These evaluations measure absolute pixel value difference and the structural fidelity of AKR signals, properties of AKR which are vital to scientists for visual analysis. Finally, we qualitatively compare DAARE to baselines when applied to real AKR signals from the South Pole Station.
Our method achieves \psnrDAARE{} PSNR and \ssimDAARE{} SSIM on synthesized AKR observations, improving PSNR by \psnrDiff and SSIM by \ssimDiff compared to baseline methods. Visual inspection of results indicate that DAARE effectively removes RFI and retains AKR structure, although spectral intensities are sometimes over-reduced. Our main contributions are as follows: (1) For RFI removal from AKR spectrograms, we present empirical evidence that stacked CDAEs achieve higher denoising performance than filtering-based and end-to-end trained deep-learning algorithms. (2) We demonstrate that employing synthetic AKR data for training is sufficient for qualitative AKR removal when applied to real observations.
\section{Background}\label{sec:background}

\subsection{Auroral radio emissions}
\ic{\ref{sec:background}.1.1 Broad related topics
\begin{enumerate}[leftmargin=1.5em]\itemsep -0.5em
    \item History of work on AKR
    \item Other auroral emissions
\end{enumerate}
}

\ai{AKR was first investigated with the satellites {IMP} 6 and 8 \cite{gurnett1974earth}, though it was probably first observed in by Elektron 2 \cite{benediktov1965preliminary, benediktov1968relation} and {OGO} 1 \cite{dunckel1970low}. The dominant theoretical model thought to generate AKR is the electron cyclotron maser instability mechanism \cite{melrose1976interpretation,wu1979theory}. 
In addition to AKR, the Earth's aurora produces four weaker radio emissions known as Auroral Hiss, Auroral Roar, and Medium-Frequency Bursts \cite{labelle2002auroral}.}

\ic{\ref{sec:background}.1.2 Importance, relevance
\begin{enumerate}[leftmargin=1.5em]\itemsep -0.5em
    \item Why is AKR important
\end{enumerate}
}

AKR is a coherent type of radio emission that can inform theoretical developments in plasma physics. The power of AKR ($\sim10^8$ watts) establishes AKR as an excellent source to study several nonlinear wave processes and wave-particle interactions \cite{gurnett1974earth}. These same interactions occur in heliospheric, planetary, and astrophysical plasmas where direct measurements may be impossible \cite{labelle2002auroral, erickson2018aero}. Investigating these processes is of great interest to forecast extreme space weather conditions and to complete our understanding of the relationships among space environments of the sun, planets, and outer solar system as well as stars and exoplanets beyond our solar system.

\ic{\ref{sec:background}.1.3 Noise sources and types of noise}
Unfortunately, observing AKR from the ground is challenging due to ionospheric refraction and absorption, as well as ground-based transmissions.
These factors have motivated the development of satellite-borne radio sensors \cite{erickson2018aero, lind2019aero} with interferometric capabilities dedicated to measuring auroral radio emissions from above the ionosphere. However, satellite missions are still affected by RFI from other sources, including solar type III bursts, planetary emissions, and space-based transmissions. A detailed description of these RFI sources are described in~\cite{desch2005quantitative}. Thus, computational noise removal can enhance the science return of satellite observations, improving examination of underlying natural processes. 

\ai{
\ic{\ref{sec:background}.1.4 Auroral kilometric radiation structure}
Previous work has demonstrated that the AKR time-frequency spectra structure is vital to proper analysis of AKR. For example, detailed comparison of AKR observations on the Geotail satellite and the South Pole Station have revealed AKR propagation towards the ground, which was previously thought to be impossible \cite{labelle2011ground}.
}

\ai{
\ic{\ref{sec:background}.1.5 Mini-review and further work
\begin{enumerate}[leftmargin=1.5em]\itemsep -0.5em
    \item Further readings
\end{enumerate}
}
Further readings for deeper understanding on auroral radiation are shared below. \cite{grabbe1981auroral} presents a full theoretical review on AKR; \cite{yearby2022review} discusses multi-spacecraft observations; and \cite{labelle2002auroral} covers non-AKR auroral radio emissions.
}

\subsection{Convolutional denoising autoencoder}

\ic{\ref{sec:background}.2.1 BM3D is (somewhat) a SOTA in denoising
\begin{enumerate}[leftmargin=1.5em]\itemsep -0.5em
    \item What is BM3D good at
\end{enumerate}
}
The BM3D algorithm \cite{dabov2007image} is considered the state-of-the-art in image denoising. However, BM3D performance decreases with images contaminated with greater amounts of noise \cite{hasan2018improved}. 

\ic{\ref{sec:background}.2.2 AutoEncoders 
\begin{enumerate}[leftmargin=1.5em]\itemsep -0.5em
    \item Where we can improve
    \item Explanation of AutoEncoders
\end{enumerate}
}
In contrast to unsupervised algorithms, denoising autoencoders have achieved similar or even improved performances for noise suppression tasks \cite{jain2008natural}.
Denoising autoencoders~\cite{vincent2008extracting} are a recent deep-learning method demonstrated to reconstruct original images from their corrupted versions across a variety of perturbations, from salt-and-pepper noise \cite{vincent2010stacked} to image masking \cite{he2022masked}. The denoising autoencoder uses a deep neural network to approximate the following:
\begin{equation}\label{eq:noise}
X = Y + Z
\end{equation}
\begin{equation}\label{eq:cdae_objective}
\hat{Y} = \psi(\phi(X))
\end{equation}
where an additive sum of the ground truth $Y$ and noise $Z$ produces a noisy observation $X$. An encoder component approximates an encoding function $\phi$ that maps the $X$ onto a latent space. The decoder component estimates a decoding function $\psi$ that produces $\hat{Y}$, a reconstruction with the minimal Euclidean distance to the ground truth $Y$.

\ic{\ref{sec:background}.2.3 AutoEncoder History
\begin{enumerate}[leftmargin=1.5em]\itemsep -0.5em
    \item History of denoising autoencoders
    \item Why AKR
\end{enumerate}
}
Many adaptations to autoencoders have augmented their ability to denoise data. Jain et al. \cite{jain2008natural} introduced convolutional layers to denoising autoencoders, achieving at par or even better than state-of-the-art denoising with other learning models. Vincent et al. \cite{vincent2010stacked} presented stacked denoising autoencoders as a preprocessing method to improve downstream classification. He et al. \cite{he2022masked} used Vision Transformers \cite{dosovitskiy2020image} for masked image inpainting and reconstruction.
These methods have been demonstrated to be robust across a variety of noise types, including complete occlusions. Their advantages identify deep-learning algorithms as a prime candidate to remove RFI present in AKR observations. 
\section{Methodology}\label{sec:methodology}
\subsection{Auroral kilometric radiation simulation}
\ic {\ref{sec:methodology}.1.1 Why use a simulation}

We explore the use of a simulation to train DAARE, since AKR data without RFI are not available. Our simulation generates a dataset of synthetic AKR spectrograms, each of $256$ by $384$ pixels dimension, to train DAARE. The dataset is drawn from a distribution of ground truth emissions $\D_Y$ and radio noise $\D_Z$. Each distribution is described by a tuple of random variables, as presented in Table~\ref{tab:simulation}, that determines sample generation. Distributions are chosen to closely imitate our sample of AKR observations from the South Pole Station. Ground truth $Y$ and noise $Z$ samples are added together to create sample observations $X$.
\begin{table}[!t]
    \renewcommand{\arraystretch}{1.3}
    \centering
    \caption{Random variables used to generate the dataset.}
    \label{tab:simulation}
    \begin{tabular}{lx{0.5\linewidth}lx{0.5\linewidth}}
        \textbf{Random variable}&\textbf{Distributed as}\\
        \toprule
        \bottomrule
        Background intensity & $U(0, 0.6)$\\
        $\text{AKR}^{(i)}$ position & $N(\theta_\text{ Width}/2,(\theta_\text{ Width}/3)^2$)\\
        & $N(\theta_\text{ Height}/2, (\theta_\text{ Height}/3)^2$)\\
        $\text{AKR}^{(i)}$ shape & $U\{0, \theta_\text{ Number of AKR shapes}\}$\\
        $\text{AKR}^{(i)}$ intensity & $N(\theta_\text{ BG intensity}, 0.15^2)$\\
        \toprule
        Gaussian noise intensity & $U(0.02, 0.08)$\\
        Overall channel intensity & $U(0.3, 0.6)$\\
        Number of channels & $Poisson(30) + 2$\\
        $\text{Channel}^{(i)}$ intensity & $U(0.1, 0.8)$\\
        \toprule
    \end{tabular}
\end{table}

\ic {\ref{sec:methodology}.1.2 Simulation random variables
\begin{enumerate}[leftmargin=1.5em]\itemsep -0.5em
    \item Ground truth features
    \item Additive noise
\end{enumerate}
}
The ground truth distribution $\D_Y$ is generated from 32 AKR shapes that include a variety of emission structures. A random background radiation intensity, several AKR shapes, and horizontal and vertical mirroring for each shape is applied.
RFI from electronics often appear in spectrograms as multiple high-intensity horizontal (nearly time-constant) emission lines in the $0-1500$ kHz frequency range, each spanning a given bandwidth of $\sim50$ kHz. A set of horizontal channels, drawn from the noise distribution $\D_Z$, is applied to resemble electronic radio emission bands. Since the Central Limit Theorem reveals that the sum of many random processes results in a Gaussian distribution, we add a layer of Gaussian noise to capture any remaining sources of noise.

\subsection{Convolutional denoising autoencoder component}
\ic {\ref{sec:methodology}.2.1 CDAE overview}
DAARE is composed of six successively stacked CDAEs with skip connections from the input observation. After the first CDAE is trained to denoise the AKR observation, the CDAE is frozen and a successive CDAE is trained on a concatenation of the first output and the original observation. The next component is trained using the same strategy. Training each component successively enables efficient parameter tuning and high denoising performance. 

\begin{figure}[!t]
\centering
\begin{minipage}[b]{\linewidth}
  \centering
  \centerline{\includegraphics[width=\textwidth]{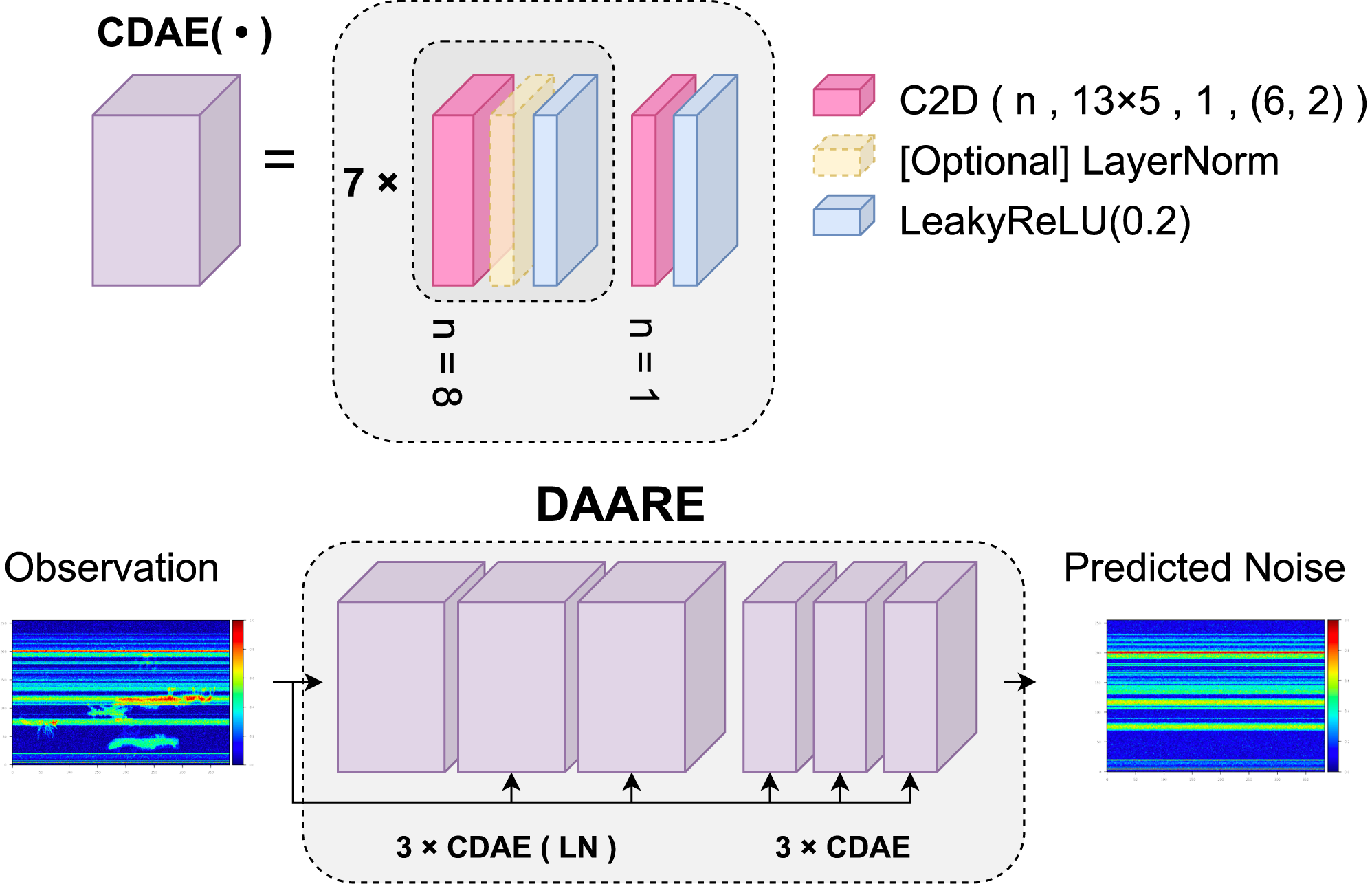}}
\end{minipage}
\caption{The DAARE architecture. C2D($n,13\times5,1,(6,2)$) stands for a $13\times5$ Conv2d layer with $n$ output channels, a stride of $1$, a padding of $(6,2)$.}
\label{fig:cdae}
\end{figure}

\ic {\ref{sec:methodology}.2.2 What is the CDAE objective}
Instead of directly reconstructing AKR, CDAE components output predictions of residual RFI, which are easier for CNNs to represent with high fidelity due to its structural simplicity. Similar approaches have been employed by prior work~\cite{zhang2016fast}. To adapt the CDAE to recognize RFI emission bands, we heighten the kernels used for convolutions to better identify thin horizontal patterns. Finally, AKR is calculated by taking the difference between the original observation and the predicted noise.
Parameters of DAARE and each CDAE component are shown in Figure \ref{fig:cdae}.

%

\subsection{Optimization}
We partition our dataset to 4096 training samples and 1024 validation samples. CDAE components are successively trained each for 10 epochs with MSE loss using a learning rate $\eta=10^{-4}$ with the Adam optimizer. We terminate the algorithm when the validation loss converges. This convergence occurs at 6 stacked components (60 epochs).

\subsection{Evaluation}
\ic{\ref{sec:results}.4.1 Evaluation
\begin{enumerate}[leftmargin=1.5em]\itemsep -0.5em
    \item Which baselines do we use
    \item Why do we use these baselines
\end{enumerate}
}
We compare DAARE against state-of-the-art filtering and deep-learning algorithms, including BF \cite{tomasi1998bilateral}, BM3D \cite{dabov2007image}, FFDNet \cite{zhang2018ffdnet}, and ViT \cite{dosovitskiy2020image}.

%
\ic{\ref{sec:results}.1.2 Baseline Optimization}
Hyperparameters for baselines are chosen to optimize PSNR. BF uses $d=32$, $\sigma_{color}=256$, $\sigma_{space}=64$. BM3D and FFDNet use a noise standard deviation $\sigma=0.2$. FFDNet, ViT, and DAARE are each trained with a total of 60 epochs at a learning rate $\eta=10^{-4}$ with the Adam optimizer.
\section{Experimental Results}\label{sec:results}

\begin{table}[!t]
    \renewcommand{\arraystretch}{1.3}
    \centering
    \setlength{\tabcolsep}{4pt}
    \caption{Experimental results of denoising methods evaluated on a test set of 1024 spectrograms.}
    \label{tab:results}
    \begin{tabular}{llllll}
        \textbf{Metric}&\textbf{BF}&\textbf{BM3D}&\textbf{FFDNet}&\textbf{ViT}&\textbf{DAARE}\\
        \toprule
        \bottomrule
        PSNR & \psnrBF & \psnrBMTD & \psnrFFDNet & \psnrViT & \psnrDAARE{b}\\
        SSIM & \ssimBF & \ssimBMTD & \ssimFFDNet & \ssimViT & \ssimDAARE{b}\\
        \# parameters & -- & -- & $486$K & $59$M & $\pmb{159\text{K}}$\\
        Avg. runtime & $\pmb{0.11\text{s}}$ & $4.63$s & $0.51$s & $0.43$s & $0.85$s\\
        \toprule
    \end{tabular}
\end{table}
\subsection{Results and analysis}

\begin{figure*}[!t]
\centering
\newcommand{\imwidth}{0.135\textwidth}
\begin{minipage}[b]{\imwidth}
  \centering
  \centerline{\includegraphics[width=\textwidth]{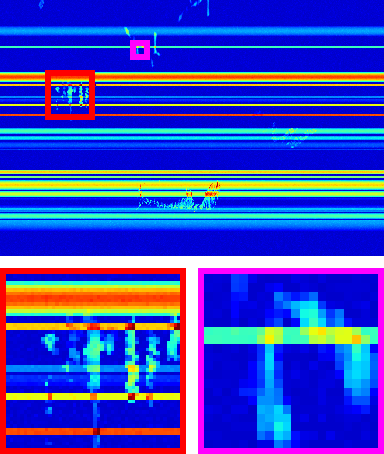}}
  \centerline{(a) Observation}\medskip
\end{minipage}
\begin{minipage}[b]{\imwidth}
  \centering
  \centerline{\includegraphics[width=\textwidth]{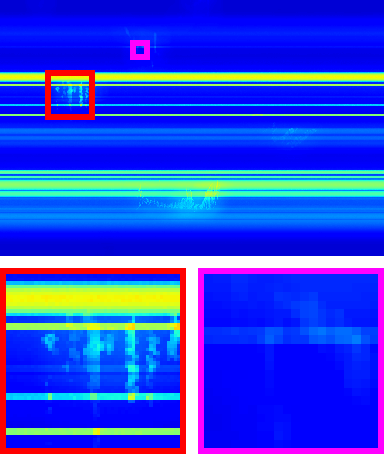}}
  \centerline{(b) BF}\medskip
\end{minipage}
\begin{minipage}[b]{\imwidth}
  \centering
  \centerline{\includegraphics[width=\textwidth]{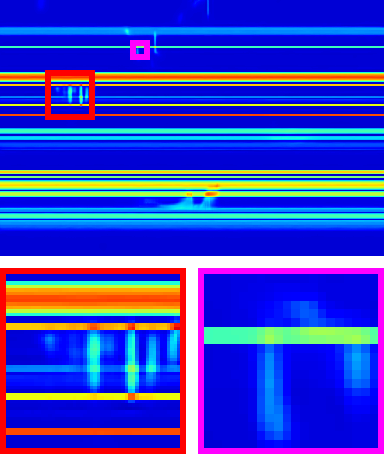}}
  \centerline{(c) BM3D}\medskip
\end{minipage}
\begin{minipage}[b]{\imwidth}
  \centering
  \centerline{\includegraphics[width=\textwidth]{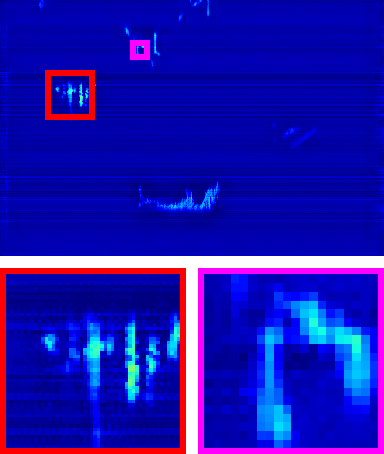}}
  \centerline{(d) FFDNet}\medskip
\end{minipage}
\begin{minipage}[b]{\imwidth}
  \centering
  \centerline{\includegraphics[width=\textwidth]{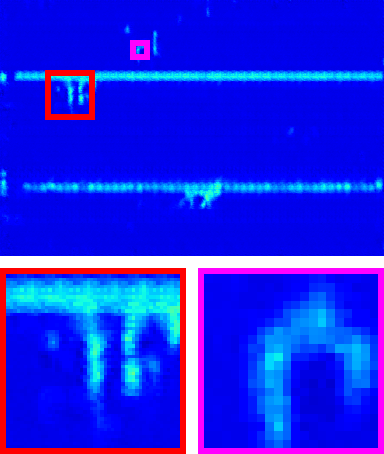}}
  \centerline{(e) ViT}\medskip
\end{minipage}
\begin{minipage}[b]{\imwidth}
  \centering
  \centerline{\includegraphics[width=\textwidth]{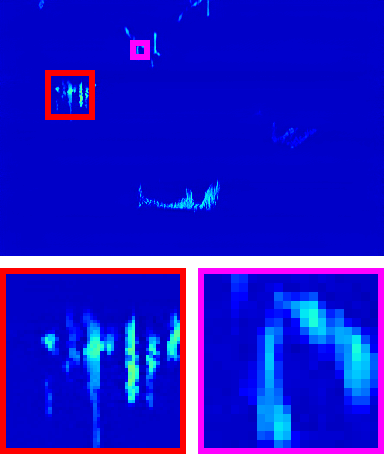}}
  \centerline{(f) DAARE}\medskip
\end{minipage}
\begin{minipage}[b]{\imwidth}
  \centering
  \centerline{\includegraphics[width=\textwidth]{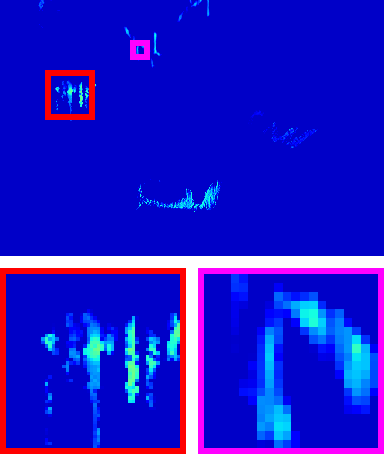}}
  \centerline{(g) Truth}\medskip
\end{minipage}
\caption{Spectrograms of denoising methods on a test observation.}
\label{fig:baselines}
\end{figure*}

\ic{\ref{sec:results}.1.1 Evaluation metrics
\begin{enumerate}[leftmargin=1.5em]\itemsep -0.5em
    \item Currently PSNR
    \item Why SSIM
\end{enumerate}
}
We evaluate results, which are presented in Table \ref{tab:results}, using a test set of 1024 simulated spectrograms with PSNR and SSIM~\cite{wang2004image}. These metrics are computed between unperturbed AKR spectrograms and the denoised AKR spectrograms. Runtime calculations for each spectrogram are performed on a one-core Intel(R) Xeon(R) CPU @ 2.20 GHz. DAARE and baseline outputs are visualized in Figure \ref{fig:baselines}.
\ai{
PSNR is commonly employed in signal processing to measure absolute pixel value reconstruction. 
However, PSNR does not always reflect human-perceived changes in structural information \cite{huynh2008scope}. For this reason, we also evaluate DAARE to baselines using SSIM to verify that reconstructed emissions closely match original AKR. 
}

\ic{\ref{sec:results}.1.2 DAARE results}
DAARE demonstrates the best denoising capability (\psnrDAARE{} PSNR, \ssimDAARE{} SSIM) on AKR, outperforming baseline denoising methods. By using multiple stacked components that successively predict residual RFI, DAARE iteratively reconstructs RFI with increasing prior intensities, resulting in an output RFI spectrogram similar to the true noise. After subtracting the predicted RFI from the input observation, we observe AKR spectrograms with high fidelity with the unperturbed AKR. As shown in Figure \ref{fig:baselines}, the DAARE output is most visually similar to the ground truth.

\ic{\ref{sec:results}.1.3 Baseline results
\begin{enumerate}[leftmargin=1.5em]\itemsep -0.5em
    \item Why baselines are bad
    \item BF, BM3D
    \item FFDNet, ViT
\end{enumerate}
}
Some limitations may contribute to a lower spectrogram denoising performance of baseline approaches.
BF and BM3D algorithms are unsupervised and are unable to recognize horizontal emission bands as noise. We observe BF to naively blur the input, which is effective for removing Gaussian noise but not RFI emission bands. Since some RFI contain sharper edges than AKR spectra, BM3D clarifies the more distinct RFI emission bands and blurs AKR. These visual observations are in agreement with low PSNR and SSIM scores in Table \ref{tab:results}.
Although FFDNet is efficient, it leaves artifacts on the resulting spectrogram, as shown in Figure \ref{fig:baselines}. This drawback outweighs its speed advantage, since scientific analysis of AKR relies on visual clarity. Similarly, while ViT has been shown to outperform CNNs on large datasets~\cite{he2022masked}, we observe that ViT outputs contain uneven pixelation when decoding the $16\times 16$ patch embeddings, resulting in its low SSIM score.

\begin{figure}[!b]
\centering

\centerline{\textbf{AKR Observations}}\medskip
\begin{minipage}[b]{.32\linewidth}
  \centering
  \centerline{\footnotesize{08-11-2016 00:40 UT}}\medskip
  \centerline{\includegraphics[width=\textwidth]{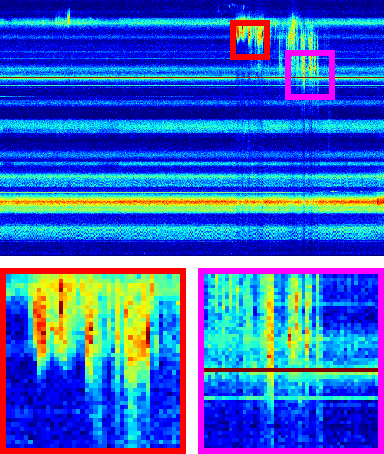}}
\end{minipage}
\hfill
\begin{minipage}[b]{0.32\linewidth}
  \centering
  \centerline{\footnotesize{07-31-2018 01:34 UT}}\medskip
  \centerline{\includegraphics[width=\textwidth]{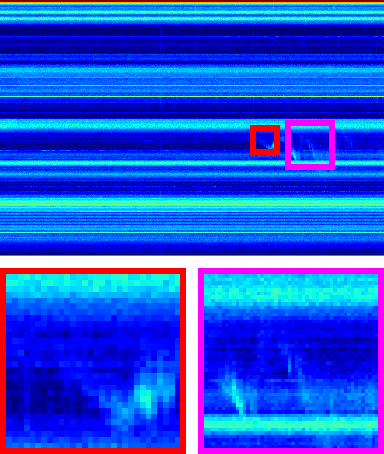}}
\end{minipage}
\hfill
\begin{minipage}[b]{0.32\linewidth}
  \centering
  \centerline{\footnotesize{07-03-2021 01:26 UT}}\medskip
  \centerline{\includegraphics[width=\textwidth]{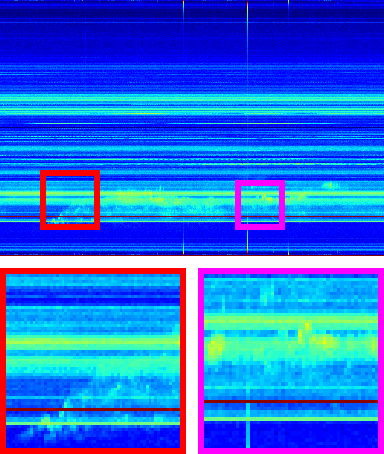}}
\end{minipage}
\\
\vspace{5pt}
\begin{minipage}[b]{\linewidth}
  \centering
  \centerline{\includegraphics[width=\textwidth]{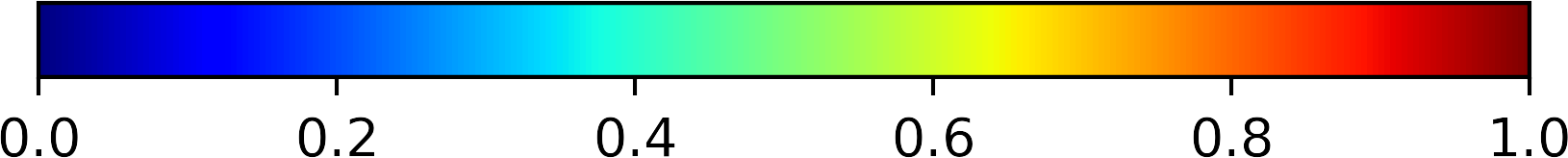}}
\end{minipage}

\centerline{\textbf{DAARE Outputs}}\medskip
\begin{minipage}[b]{.32\linewidth}
  \centering
  \centerline{\footnotesize{08-11-2016 00:40 UT}}\medskip
  \centerline{\includegraphics[width=\textwidth]{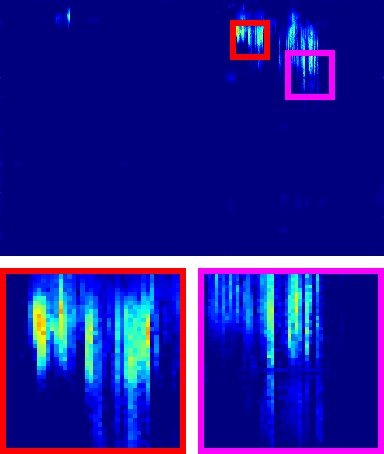}}
\end{minipage}
\hfill
\begin{minipage}[b]{0.32\linewidth}
  \centering
  \centerline{\footnotesize{07-31-2018 01:34 UT}}\medskip
  \centerline{\includegraphics[width=\textwidth]{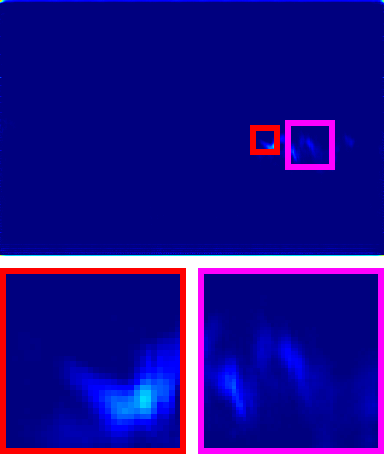}}
\end{minipage}
\hfill
\begin{minipage}[b]{0.32\linewidth}
  \centering
  \centerline{\footnotesize{07-03-2021 01:26 UT}}\medskip
  \centerline{\includegraphics[width=\textwidth]{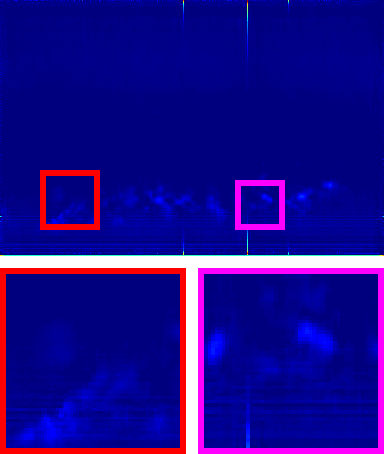}}
\end{minipage}
\\
\vspace{5pt}
\begin{minipage}[b]{\linewidth}
  \centering
  \centerline{\includegraphics[width=\textwidth]{jet_cbar}}
\end{minipage}
\caption{AKR observations and DAARE outputs on a normalized $(\mu=0.2, \sigma=0.2)$ dB scale from the South Pole Station.}
\label{fig:real_obs}
\end{figure}
\subsection{Inspections on real observations}
\ic{\ref{sec:results}.2.1 Real observation results
\begin{enumerate}[leftmargin=1.5em]\itemsep -0.5em
    \item Real observation figures
    \item Qualitative analysis of figures
\end{enumerate}
}
As shown in Figure \ref{fig:real_obs}, DAARE outputs share high structural fidelity with AKR observations. The magnifications in ``08-11-2016 UT" and ``07-31-2018 UT" share nearly identical structure and intensity without the interference of Gaussian noise or horizontal emission bands. High structural fidelity enables scientists to uncover and evaluate properties of AKR without ground-level noise disturbances. These comparisons indicate that DAARE is able to effectively remove RFI from AKR spectrograms despite being trained with synthetic data. However, as shown in ``07-03-2021 UT," removing RFI has an unintended effect of dampening the intensity of spectral features. These differences may affect implications from downstream analysis. Further refinement of synthetic data may result in greater conservation of spectral intensities in deep-learning denoisers.
\section{Conclusion and Future Work}\label{sec:conclusion}
\ic{\ref{sec:conclusion}.1 Conclusion:}
We show that deep learning tools can be used to enhance image representations of astronomy data by removing visual noise. Stacked CDAEs outperform baselines for RFI removal from a simulated AKR test bed. We achieve this with DAARE, a model that is trained with synthetic data but is still able to effectively remove RFI from real AKR data from the South Pole Station. 
DAARE can pick out AKR in time-frequency spectrograms and remove strong horizontal RFI emission bands. 
Modifying the simulated data allows DAARE to be generalized to treat other structures of RFI.

\ic{\ref{sec:conclusion}.2 Downstream applications:
\begin{enumerate}[leftmargin=1.5em]\itemsep -0.5em
    \item Ground-level observations to spacecraft-level observation translation
    \item Automatic recognition and classification of AR
\end{enumerate}
}
Our results have implications for space science analysis beyond this study. In addition to cleaning AKR for scientists, removing RFI is a crucial preprocessing step for analysis of spectrographic data, a visualization method of radio frequency data common to many works in radio astronomy. Future work can leverage denoised spectrograms for a plethora of algorithmic science analysis, including automatic detection and forecasting of natural auroral events, emission clustering, or data comparison across long geographic distances.
\section{Acknowledgements}

This work was supported by National Science Foundation grant AST-1950348. The authors acknowledge the MIT SuperCloud and Lincoln Laboratory Supercomputing Center for providing (HPC, database, consultation) resources that have contributed to the research results reported within this paper. 
%

\vfill\pagebreak
\bibliographystyle{IEEEbib}
\bibliography{refs/aurora, refs/autoencoders, refs/evaluation, 
refs/similarpapers, refs/template_references}


\end{document}